\documentclass[twocolumn,prb,showpacs,preprintnumbers,amsmath,amssymb]{revtex4}
\usepackage{graphicx}
\usepackage{dcolumn}
\usepackage{bm}

\begin{document}
\preprint{Chalmers Applied Physics Reports 2001--61}
\title{Theory for structure and bulk-modulus determination}
\author{Eleni Ziambaras}%
\author{Elsebeth Schr\"oder}\thanks{Corresponding author}%
\email{schroder@fy.chalmers.se}
\affiliation{Applied Physics, Chalmers University of
Technology and G\"oteborg University, SE--41296 Gothenburg, Sweden}
\date{February 22, 2003}

\begin{abstract}
A new method for direct evaluation of both crystalline
structure, bulk modulus $B_0$, and bulk-modulus 
pressure derivative $B'_0$ of solid materials with complex crystal structures
is presented.  The explicit and exact 
results presented here permit a multidimensional polynomial fit of the total 
energy as a function of all relevant structure 
parameters to simultaneously determine the equilibrium 
configuration and the elastic properties.  The method allows 
for inclusion of general (internal) structure parameters, \textit{e.g.}, 
bond lengths and angles within the unit cell, on an equal footing 
with the unit-cell lattice parameters.  The method is illustrated 
by the calculation of $B_0$ and $B'_0$ for a few selected 
materials  with multiple structure parameters  for 
which data is obtained by using first-principles density functional 
theory. 
\end{abstract}


\maketitle

\section{Introduction}

Calculations of the bulk structure and the bulk elastic
properties play an important role in the physics of condensed 
matter.\cite{alchagirov,kackell,persson,kappaalumina,pre00,wach}
Bulk calculations help us understand, characterize, and 
predict mechanical properties of materials in our 
surroundings, under extreme conditions, as in geological 
formations and setting,\cite{katahara} and for industrial 
applications.\cite{kappaalumina,japan} 
Crystalline materials come in many different structures and, in contrast
to isotropic materials,  the structure description of crystalline materials 
may in general need multiple lattice parameters and an atomic basis. In 
this paper we discuss how to determine the equilibrium structure 
of a (multiparameter) crystalline material while, at the same time, directly 
determining the bulk modulus and the bulk modulus pressure derivative. We 
argue and show that for theoretical structure calculations of multiparameter 
systems this is simpler and more exact than fitting to (semi-)empirical 
equations of state (EOS) such as, \textit{e.g}, the Murnaghan or Birch EOS. 
In particular, with our direct method there is no need to first determine 
the hydrostatic path of the system. We further discuss how to include the 
atomic basis in this process in a natural way.

In crystalline materials described by a single lattice parameter (\textit{e.g}, 
monatomic cubic phases) the lattice parameter is a simple function of the 
unit-cell volume, and the equilibrium volume thus uniquely determines the 
equilibrium structure, \textit{i.e.}, the value of the lattice parameter.
This is not the case when multiple lattice parameters characterize the 
system and a whole range of lattice-parameter values can form the 
same unit-cell volume.
The equilibrium structure of the material must then be found
by fitting and minimizing the free energy within the 
multidimensional space of lattice parameters.\cite{kackell,wach,comment}
Relevant variables describing the atomic basis (\textit{e.g.}, bond 
lengths or binding angles) may be included
among the parameters, and the full set of lattice parameters and internal 
(atomic basis) parameters are collected into the vector $\bm{x}$, 
scaled to dimensionless form. The volume of the unit cell $V(\bm{x})$ 
depends in a simple way
on the values of the lattice parameters describing the unit cell, 
but not on the internal atomic configuration. Nevertheless, we here 
treat the external and internal parameters on an equal footing.

{}From theory bulk calculations the total energy 
(per unit cell) $E(\bm{x})$ is found for a number of 
structures $\bm{x}$.
The elastic response of typical hard 
crystalline materials corresponds to small deviations
$\delta \bm{x} =\bm{x}- \bm{x}^{(0)}$ of the structural 
parameters from the equilibrium structure $\bm{x}^{(0)}$.
The observation that the 
total energy forms a natural potential (hyper-)surface in the parameter 
space of lattice and internal parameters $\bm{x}$, combined with
the accuracy of present-day bulk-calculation methods
(such as density functional theory, embedded atom 
methods, or effective medium theory),
then makes it 
possible to fit the corresponding total-energy variation 
through the multidimensional fit
\begin{equation}
E(\bm{x})= k+\frac{1}{2}M_{ij} \delta x_i\delta x_j 
+\frac{1}{3!} \gamma_{i j l}\delta x_{i} \delta x_{j}\delta x_{l}
+ O\left( \delta\bm{x} \right)^4\, ,
\label{eq:mpfit}
\end{equation} 
at controlled accuracy.
Here $k$, $M$, and $\gamma$ denote zeroth, second, and 
third-rank tensors of fitting constants. An additional set of fitting
constants are the $\bm{x}^{(0)}$ hidden in 
$\delta \bm{x} =\bm{x}- \bm{x}^{(0)}$
 The polynomial 
fit~(\ref{eq:mpfit}) gives a transparent description of 
the materials-structure energy variation and directly determines
the equilibrium structure $\bm{x}^{(0)}$. 

In this paper we exploit and use the structure
calculation, \textit{i.e.}, the multidimensional polynomial 
fit of the total energy (\ref{eq:mpfit}) for an
additional and direct determination of the zero-pressure 
bulk modulus $B_0=-V(\bm{x}^{(0)}) (\partial p /\partial V) 
|_{\bm{x}=\bm{x}^{(0)}}$ and its pressure 
derivative, 
$B'_0=-\partial \{V (\partial p /\partial V) \}/\partial p
|_{\bm{x}=\bm{x}^{(0)}}$ at zero temperature.

For a general set of structure parameters, $\bm{x}$,
we expand the volume around the equilibrium configuration 
$\bm{x}^{(0)}$ using the gradient $\bm{g}=\left.\nabla 
V(\bm{x}) \right|_{\bm{x} =\bm{x}^{(0)}}$ and 
the Hessian
$H=\left.H \left( V(\bm{x})\right) \right|_{\bm{x}=\bm{x}^{(0)}}=
\left[ \left\{\partial^2 V(\bm{x})/(\partial x_i\partial x_j)
\right\}_{ij}\right]_{\bm{x}=\bm{x}^{(0)}}$ of the volume. We note that
derivatives of the volume with respect to the internal
parameters vanish, by definition. 
By providing a systematic treatment of the  
structural changes induced by the pressure $p=-\partial E/\partial V$
we extract from the minimum of the 
zero-temperature enthalpy
\begin{equation}
{\cal H}(\bm{x},p)=E(\bm{x}) + p V(\bm{x})
\label{eq:enthalpy}
\end{equation}
both the bulk modulus
\begin{equation}
B_0
= \frac{V(\bm{x}^{(0)})}{\bm{g}^T M^{-1}\bm{g}}
\label{eq:mpbulk}
\end{equation}
and the bulk-modulus pressure derivative
\begin{widetext}
\begin{equation}
B'_0=
V(\bm{x}^{(0)}) \frac{3\bm{g}^T M^{-1}H M^{-1}\bm{g}-
\gamma_{i j l} 
\left(M^{-1}\bm{g}\right)_{i}
\left(M^{-1}\bm{g}\right)_{j}
\left(M^{-1}\bm{g}\right)_{l}}%
{\Big( \bm{g}^T M^{-1} \bm{g}\Big)^2}-1.
\label{eq:mpbulkderiv}
\end{equation}
\end{widetext}
The algorithm outlined above can also be applied to the corresponding 
direct determinations of general harmonic\cite{nye} and anharmonic
elastic properties.\cite{Forthcoming} 

Our results both enhance the theory understanding 
of the crystalline mechanical properties and simplify the
desired testing of theory calculations as they combine  
the formal determination of the crystalline structure 
[Eq.~(\ref{eq:mpfit})] and of the elastic properties 
[Eqs.~(\ref{eq:mpbulk}) and (\ref{eq:mpbulkderiv})]. 
For example, from Eqs.~(\ref{eq:mpfit}) and 
(\ref{eq:enthalpy}), we can directly identify which 
(internal) structure parameters softens the bulk 
modulus~(\ref{eq:mpbulk}) and we may, in turn, strengthen 
the materials by suitable chemical or structural modification.  

For cases like graphite, where the state-of-the-art DFT based on 
the generalized gradient approximation (GGA)
fails to describe the weak physical interlayer 
binding, the polynomial fit~(\ref{eq:mpfit}) 
highlights the intrinsic theory challenges:\cite{japan}
the total energy has no minimum in the space of
lattice parameters, the too soft dependence of the total energy
on the graphite-layer separation is quite apparent in the fit.
At the same time, for strongly bonded crystalline 
materials the fit~(\ref{eq:mpfit}) produces usable estimates of 
the equilibrium crystalline structure and shows the strength of 
GGA-DFT. The equilibrium volume, the crystallographic parameters,
and the bulk modulus, describing the material's resistance
to hydrostatic stress,
provide simple experimental tests against which we can 
compare and calibrate our calculations.  

Besides the direct relevance of our results for the description of 
complex materials our calculations of bulk structure and bulk modulus 
calculations are also of interest for development of 
pseudo-potential-based density-functional-theory (DFT) methods and 
for methods using empirical parameters. There, a first and critical 
test of the pseudopotential or the empirical parameters  
is whether the calculations predict a correct materials 
structure, binding, and elastic properties for the relevant equilibrium 
configuration. Present DFT scripts\cite{dacapo} can 
automate some pseudopotential testing for simple materials and symmetries,
our formal results generalize such testing 
of theory accuracy to cases when multiple structural
parameters determine the elastic properties.

The outline of this paper is as follows.
In Section II we discuss the traditional methods of determining the
bulk modulus for single- and multiparameter systems. In Section III 
we derive our expressions for the bulk modulus and the bulk modulus derivative, 
Eqs.~(\ref{eq:mpbulk}) and (\ref{eq:mpbulkderiv}), for the
simple one-parameter problem ({\em e.g.}, mono-atomic fcc or bcc structures), 
easily generalized to the $n$-parameter problem.
In Section IV
we proceed to illustrate and test the algorithm
on a number of mono- and di-atomic materials based on first-principle 
DFT calculations and comparison to experiments. Comparisons
of $B_0$ and $B'_0$, together with the test of the lattice and structure 
parameters themselves, represent the typical test of 
materials-theory accuracy. Section V contains the conclusion.

\section{Background}

A theory determination of the zero-temperature bulk modulus based on
either traditional methods\cite{alchagirov,murnaghan,birch}
or our formal result~(\ref{eq:mpbulk}) is straightforward 
when one single structural parameter (\textit{e.g.,} the
lattice parameter $a$) defines the crystalline state.
This situation applies for monatomic crystals with simple cubic (sc), 
face-centered cubic (fcc) and body-centered cubic (bcc)
symmetries.  Here, the
unit-cell volume $V(a)=qa^3$ uniquely determines the lattice
parameter $a$ through a dimensionless number $q$ which depends on the
crystal symmetry ($q=1$, $q=1/4$, and $q=1/2$ for sc, fcc, and bcc lattices, 
respectively). All which is required are theory 
calculations of total energies for a range of $a$ values to 
determine both the equilibrium structure $a_0$ and the
equilibrium volume $V^{(0)}$. The total energy per unit 
cell, $E(a)$ (as in Eq.~(\ref{eq:mpfit})), can then be 
expressed as a function of the unit-cell volume, $E(V)$.

\begin{figure}[hl]
\begin{center}
\scalebox{0.96}{\includegraphics{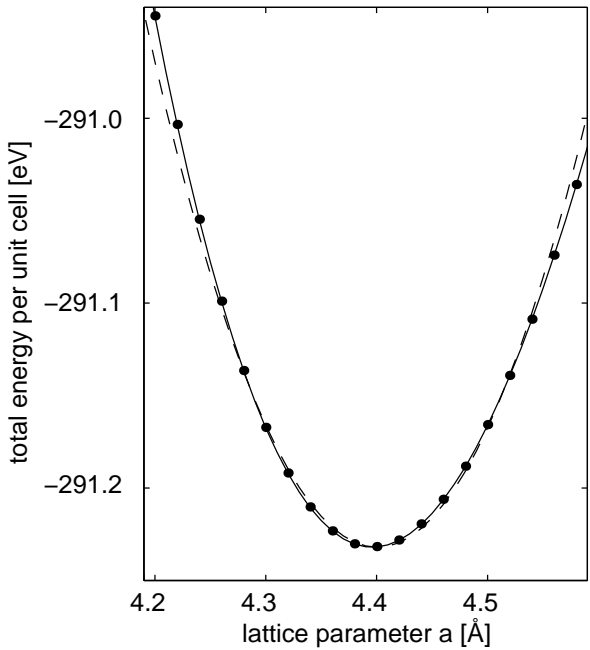}}
\caption{\label{figetot1d}The total energy per unit cell (two atoms) as 
a function of the lattice 
parameter $a$ for the 3C polytype of SiC. The circles are the data points
obtained from DFT calculations. {\em Solid line:} Fourth-order polynomial
fit as used for the values in Table~\protect\ref{table1}. 
{\em Dashed line:} Second-order
polynomial fit to the 15 central data points.}
\end{center}
\end{figure}

The general approach is illustrated by the example in Figure~\ref{figetot1d},
which shows the total energy as a function of the lattice 
parameter $a$ for the zinc-blende phase of SiC (3C-SiC),
as found from DFT calculations. A parabola fit to the 15 data points
closest to the equilibrium value of $a$ and a fourth-order polynomial fit 
to all data points are shown.
Fits using the traditional 
Murnaghan equation of state,\cite{murnaghan} integrated
to give
\begin{equation}
E_{\text{Murn}}(V)=-E_0+
\frac{B_0V}{B'_0}
\left[\frac{(V^{(0)}/V)^{B'_0}}{B'_0-1}+1\right]
-\frac{V^{(0)}B_0}{B'_0-1}
\label{eqMurn}
\end{equation}
or the Birch equation of state,\cite{birch} integrated to
\begin{eqnarray}
E_{\text{Bir}}(V)
&=&-E_0+
\frac{9}{8}B_0V^{(0)}
\left[ \left(V^{(0)}/V\right)^{2/3}-1\right]^2
\nonumber \\[0.6em]&&
+\frac{9}{16}B_0V^{(0)}
\left\{ B'_0-4\right\}
\left[ \left(V^{(0)}/V\right)^{2/3}-1\right]^3
\nonumber \\[0.6em]&&
+{\cal O}\left[ \left(V^{(0)}/V\right)^{2/3}-1\right]^4,
\label{eqBir}
\end{eqnarray}
yield, to the eye, curves identical to the fourth-order polynomial fit and 
are not shown separately. In Murnaghan's and Birch's equations 
(\ref{eqMurn}) and (\ref{eqBir}) the quantities $B_0$, $B'_0$, 
and $V^{(0)}$, and in some cases also the cohesive energy $E_0$,
are fitted. Other equations of state traditionally used are mentioned in 
Refs.~\onlinecite{alchagirov} and \onlinecite{birch}.

The values of the equilibrium lattice parameter $a_0$, and of 
$B_0$ and $B'_0$ obtained from Eqs.~(\ref{eq:mpbulk}) and 
(\ref{eq:mpbulkderiv}) and 
from the Murnaghan and Birch fits are included 
in Table~\ref{table1}. For systems described by one lattice parameter
Eqs.~(\ref{eqMurn}) and (\ref{eqBir}) give bulk moduli
and bulk-modulus derivatives in close agreement with our present
direct approach, Eqs.~(\ref{eq:mpbulk}) and (\ref{eq:mpbulkderiv}).

We would like to stress that the moduli $B_0$ and $B'_0$ are formally 
defined as zero-pressure quantities, and in no way depend on 
finite-pressure behavior 
beyond the pressure gradient at $p=0$. If we are able to sample our
theory system in a sufficiently dense grid around the zero-pressure structure
the values of $V_0$, $B_0$, and $B'_0$ in~(\ref{eq:mpbulk}) 
and~(\ref{eq:mpbulkderiv}) are exact,\cite{accuracy} and can be related to the
corresponding exact determination of the elastic constants.
Fits to empirical EOS may yield results of $V_0$, $B_0$, and $B'_0$ that are 
in good agreement with experimental observations, but they do not 
necessarily constitute the exact quadratic response.

\begin{table*}
\caption{\label{table1}
Bulk properties calculated from DFT data obtained directly from
Eqs.~(\ref{eq:mpbulk}) and (\ref{eq:mpbulkderiv}) via a fourth-order
polynomial fit
(``Present approach''), available experimental values, and values from fits to 
Murnaghan's and Birch's equations (\ref{eqMurn}) and (\ref{eqBir})
along the hydrostatic path.
For the internal parameter in 2H-SiC we find the following value:
Si-C distance along the $c$-direction $u$(Si-C)$=0.3752c$ or 
bondlength $\ell_{bond}=1.9031$ \AA.
(Experiments find\protect\cite{schulz} $u$(Si-C)$_{\text{exp}}=0.3760c$,
${(\ell_{\rm bond})}_{\rm exp}=1.8998$ \AA. )
For 4H-SiC we find ${u({\rm Si-C})}_1=0.1880c$, ${u({\rm Si-C})}_2=0.1874c$, 
$u{(\rm Si-Si)}=0.2500c$, in good agreement with other theoretical 
results.\cite{kackell, bauer}}

\begin{ruledtabular}
\begin{tabular}{llcccccccccccc}
 &&\multicolumn{4}{c}{Present approach} & \multicolumn{4}{c}{Experiment} & 
\multicolumn{2}{c}{Murnaghan} & \multicolumn{2}{c}{Birch} \\

 && $a_0$ [\AA] & $c/a$ & $B_0$ [GPa] & $B'_0$ & $a_0$ [\AA]& $c/a$ & 
$B_0$ [GPa] & $B'_0$&  $B_0$ [GPa] & $B'_0$&  $B_0$ [GPa] & $B'_0$ \\
\hline
Co & fcc & 3.531 &       & 218 & 4.80 & && & & 214 & 4.47 & 216 & 4.65 \\
   & bcc & 2.817 &       & 206 & 5.09 & && & & 199.4 & 4.90 & 205 & 4.95 \\
   & hcp & 2.500 & 1.617 & 223 & 4.70 & 2.51\footnotemark[1] &1.622\footnotemark[1] & 191.4\footnotemark[1] &  5.07\footnotemark[2] & 217& 4.52& 219& 4.61\\
\hline %
SiC & 3C & 4.376 &       & 213 & 3.93 & 4.3596\footnotemark[3] && 224\footnotemark[4] & 4.0\footnotemark[5] & 212 & 3.87 & 213 & 3.91 \\ 
& 2H  & 3.089 & 1.642 & 213 & 3.92 & 3.079\footnotemark[6] &1.641\footnotemark[6]& 223\footnotemark[7] & & 208 & 3.74 & 211 & 3.86 \\    
 & 4H  & 3.092 & 3.274 & 213 & 3.93 & 3.073\footnotemark[3]& 3.271\footnotemark[3] &&& 212 & 3.78 & 213 & 3.89 \\  
\hline
\hline
C  & diam& 3.565 &       & 436 & 3.71 & 3.567\footnotemark[1] && 443\footnotemark[1] & 4.07\footnotemark[8]  & 432 & 3.72 & 435 & 3.70 \\
\hline
Si & diam& 5.466 &       & 88.7 & 4.35 & 5.430\footnotemark[1] && 98.8\footnotemark[1] & 4.09\footnotemark[9] & 87.7 & 4.20 & 88.3 & 4.28 \\
\end{tabular}
\end{ruledtabular}
\footnotetext[1]{Ref.~\onlinecite{kittel}}
\footnotetext[2]{Ref.~\onlinecite{batsanov}}
\footnotetext[3]{Ref.~\onlinecite{landolt17c}}
\footnotetext[4]{Ref.~\onlinecite{landolt22a}}
\footnotetext[5]{Ref.~\onlinecite{strossner}}
\footnotetext[6]{Ref.~\onlinecite{schulz}}
\footnotetext[7]{Ref.~\onlinecite{carnaham}}
\footnotetext[8]{Ref.~\onlinecite{mcskimin}}
\footnotetext[9]{Ref.~\onlinecite{beattie}}
\end{table*}

For materials with multiple structure parameters, 
the procedure of the traditional approaches further becomes quite 
awkward as it must be supplemented by a separate discussion of how the
experimental conditions define the relevant structural constraint at 
a given volume, the hydrostatic path $\bm{x}=\bm{x}(V)$.
Moreover, cross-correlations on the $n$-dimensional
energy surface are ignored in traditional fitting procedures.
These procedures are basically a one-dimensional fit in the
$n$-dimensional space and they are thus more subject to
numerical noise in the data points than our approaches based
on the multidimensional least-squares polynomial fit 
(\ref{eq:mpfit}).\cite{kackell}

\begin{figure}
\begin{center}
\scalebox{0.8}{\includegraphics{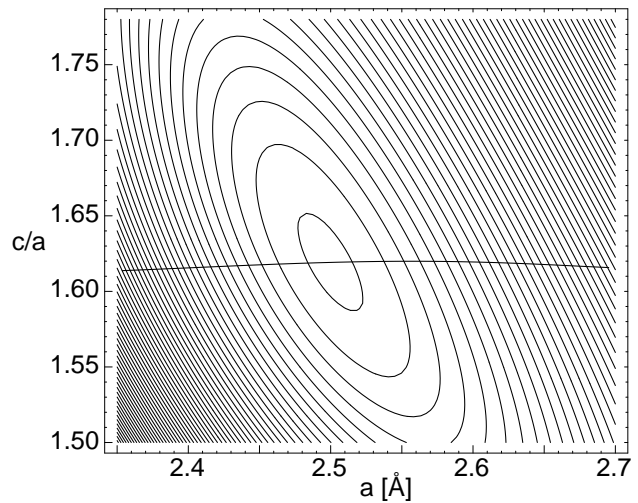}}
\caption{\label{figCohcp}Contour plot of the fourth-order polynomial
fit to the Co hcp total energy, including the hydrostatic path.
The hcp unit cell is given by the two independent lattice parameters
$a$ and $c/a$. The contour step is $0.025$ eV per unit cell (two atoms).}
\end{center}
\end{figure}

A simple multi-parameter case illustrates this point.
Figure~\ref{figCohcp} both describes the energy surface, 
Eq.~(\ref{eq:mpfit}), fitted through DFT calculations for Co, 
and emphasizes the general advantages of our direct 
approach [Eqs.~(\ref{eq:mpbulk}) and (\ref{eq:mpbulkderiv})] 
compared with the traditional bulk-modulus 
determinations.\cite{alchagirov,murnaghan,birch}
Materials like Co, which has a non-ideal hexagonal close-packed (hcp) 
structure, graphite with its layered structure, or the 
polytypes of SiC\cite{kackell} or alumina,\cite{kappaalumina} have 
multiple lattice parameters 
(plus relevant internal degrees of freedom) which are, of course, 
no longer uniquely specified by the volume but depend on the 
general materials conditions. The hydrostatic path 
defines the system when subject to uniform pressure (as 
relevant for the bulk-modulus measurements). The hydrostatic 
path is identified in Fig.~\ref{figCohcp}.
The traditional approaches\cite{murnaghan,birch} 
proceed by implementing this complex constraint in the
equation of state, Eq.~(\ref{eqMurn}) or (\ref{eqBir}), 
to fit the bulk modulus and its derivative. Instead, we present an 
explicit determination, Eqs.~(\ref{eq:mpbulk}) and
(\ref{eq:mpbulkderiv}), based directly on the
equation of state (\ref{eq:mpfit}) expressed
as a function of the underlying crystalline structural parameters.

\section{Derivation}

Our direct bulk-modulus evaluations, Eqs.~(\ref{eq:mpbulk}) 
and (\ref{eq:mpbulkderiv}), are the results of using the pressure 
(instead of the volume) in a formal identification of the 
hydrostatic path and then invoking a systematic expansion in small pressure for
an explicit specification of $B_0$ and $B'_0$.  
Today DFT and other materials-theory bulk-calculations are done with 
high accuracy, and we need only to vary the
lattice parameters values slightly around the optimal structure
to approximate the total-energy
curve by an accurate polynomial fit, Eq.~(\ref{eq:mpfit}). The minimum of the 
corresponding (zero-temperature) enthalpy~(\ref{eq:enthalpy}) 
can thus be used to directly specify the physically correct 
structural configuration at any given pressure $p$. The set 
of these optimal structure-parameter values, $\bm{x}_{\text{hydro}}
(p)$, trace out the hydrostatic path which, when parameterized by 
$p$, is obtained by simply solving the equation
\begin{equation}
\nabla {\cal H}(\bm{x},p) = \nabla E(\bm{x}) + p \nabla V(\bm{x}) =0.
\label{eqnablaH}
\end{equation}
We obtain a formal expression for the general (pressure dependent) bulk modulus 
by taking the derivative along this hydrostatic path
\begin{equation}
B = -V(\bm{x}_{\text{hydro}}(p))
\left(\frac{\partial V(\bm{x}_{\text{hydro}}(p))} 
{\partial p}\right)^{-1}, 
\end{equation}
and finally we extract the explicit results for the
zero-pressure bulk-modulus $B_0$ and for its pressure 
derivative, $B'_0$.

We illustrate the general derivation by focusing on a one-dimensional
parameter space $x$, {\em e.g.}, for the fcc or bcc
one-atomic structure. Then the total energy can be fitted by the polynomial
\begin{equation}
E(x)= k+\frac{1}{2}  M (x-x^{(0)})^2 + \frac{1}{3!}\gamma(x-x^{(0)})^3+
f(x-x^{(0)})
\label{eq1dimfit}
\end{equation} 
where $f(x-x^{(0)})={\cal O}(x-x^{(0)})^4$ contains higher order terms. 
The coefficients $k$, $M$, $\gamma$, the coefficients of 
$f(x-x^{(0)})$, as well as the optimal value of the lattice 
parameter at zero pressure, $x^{(0)}$, are the fitting parameters
to be specified, for example,  by a set of accurate underlying DFT
calculations.  

At small pressures, {\em i.e.}, for lattice and internal parameter values 
close to the zero-pressure optimal values, the $pV$ term in the enthalpy
is small and can be regarded as a perturbation of the system.
To proceed we introduce a small, non-dimensional and real 
parameter $\lambda$ such that we can write the small pressure 
as $p=\lambda p^{(1)}$ and the lattice-parameter variable as 
$x=x^{(0)}+\lambda x^{(1)}+\lambda^2 x^{(2)}+{\cal O}(\lambda^3)$. 
The variables $x^{(1)}$, $x^{(2)}$, \ldots\ are unknown, are functions of the 
pressure and must be found in the following. 
 
The bulk modulus expression requires calculation of the volume $V(x)$ and its 
pressure derivative. We write the volume in a Taylor-expansion 
around the zero-pressure solution $x^{(0)}$ as
\begin{eqnarray}
V(x)&=&V(x^{(0)}) + \lambda g x^{(1)}
\nonumber \\[0.6em]
&&
+\lambda^2\left( x^{(2)}g+\frac{1}{2!} \left(x^{(1)}\right)^2H \right) 
+{\cal O}(\lambda^3)
\label{eqvoltaylor}
\end{eqnarray}
with
$g=\left.d V/dx\right|_{x=x^{(0)}}$
and 
$H=\left. d^2 V/d x^2\right|_{x=x^{(0)}}$.
Here, the 
pressure dependence enters through the variables
$x^{(1)}$, $x^{(2)}$, \ldots . The pressure derivative of
the volume is thus 
\begin{equation}
\frac{\partial V(x)}{\partial p}
= g\frac{\partial x^{(1)}}{\partial p^{(1)}} 
+\lambda\left( g
\frac{\partial x^{(2)}}{\partial p^{(1)}} 
+ \frac{\partial x^{(1)}}{\partial p^{(1)}} H x^{(1)}
\right)
+{\cal O}(\lambda^2)
\label{eqbulksemi}
\end{equation}
and we determine the variables
$x^{(1)}$, $x^{(2)}$, \ldots\ by solving the condition on
the enthalpy given by (\ref{eqnablaH})
\begin{eqnarray}
0&=&
\lambda\left( M x^{(1)}+p^{(1)} g \right)
\nonumber\\[0.6em]
&&+ \lambda^2 \left( M x^{(2)}+\frac{1}{2}\gamma\left(x^{(1)}\right)^2
+p^{(1)}H x^{(1)}\right)+{\cal O}(\lambda^3)\,.
\nonumber\\ \label{eq:idlambda}
\end{eqnarray}
The identity (\ref{eq:idlambda}) must hold for every order and we thus 
obtain a formal pressure dependence of the lattice parameter
\begin{eqnarray}
x^{(1)}
&=&
-p^{(1)} M^{-1}g
\\[0.6em]
x^{(2)}
&=& \left({p^{(1)}}\right)^2 \left\{ M^{-1}H M^{-1} g
-\frac{1}{2}M^{-1}\gamma M^{-1} g M^{-1} g \right\} \,.
\nonumber\\
\end{eqnarray}

Finally, introducing these solutions into (\ref{eqbulksemi}) we find 
for $\lambda=0$ the isothermal zero-pressure bulk modulus
\begin{equation}
B_0=-V(x^{(0)}) 
\left(\frac{\partial V}{\partial p}\right)_{x=x^{(0)}}^{-1}
= \frac{V(x^{(0)})}{g M^{-1}g}
\end{equation}
and taking the derivative of 
$-V \left(\partial V/\partial p\right)^{-1}$
with respect to $p=\lambda p^{(1)}$ we find
at $\lambda=0$ 
\begin{equation}
B'_0=
V(x^{(0)}) \frac{3g M^{-1}H M^{-1}g-\gamma M^{-1} g M^{-1}
g M^{-1} g}%
{\left( g M^{-1} g\right)^2}-1
\end{equation}
in the case when one (lattice) parameter suffices to describe the unit cell
and its atom basis. 

The above derivation is straightforwardly generalized to
materials systems in which $n$ independent lattice and internal 
parameters determine the structure and the bulk moduli
Eqs.~(\ref{eq:mpbulk}) and (\ref{eq:mpbulkderiv}), which is our main result.
We stress that $B_0$ and $B'_0$ are evaluated at zero pressure and thus 
the results are exact in spite of the perturbation.
$B_0$ and $B'_0$ depend directly on the second order, respectively on 
the second and third order, coefficients of the energy fit ($M$ and 
$\gamma$). We observe that the coefficients of 
$f(x-x^{(0)})$ do not enter the expression for the bulk modulus 
(\ref{eq:mpbulk}) or the pressure derivative of the bulk modulus 
(\ref{eq:mpbulkderiv}). However, their 
presence may improve the fit (\ref{eq1dimfit}), and thereby 
affect also the coefficients $M$, and $\gamma$, 
and thus $B_0$ and $B'_0$. Internal parameters,
which describe the positions of the atoms within the unit cell, naturally 
do not enter the expression of the volume, and thus not the volume derivatives
$\bm{g}$ and $H$ either, but do affect $B_0$  and  $B'_0$ through $M^{-1}$. 

Higher pressure derivatives of the bulk modulus
may be found by taking into account the higher orders of $\lambda$ in the
Taylor expansions of $\bm{x}$ and the volume. The pressure derivatives will 
depend on successively higher orders in the polynomial fit. The derivation is 
straightforward if somewhat tedious.

\section{Examples of applications}

As an example of the use of the algorithm for determining $B_0$ and $B'_0$
we evaluate the structure and
bulk modulus of a selection of one and two-species materials. We fit data
obtained from DFT calculations, described in further detail below,
to fourth-order polynomials of the form 
(\ref{eq:mpfit}) in $n$-dimensional space, where $n=1$, 2, 3, or 5. 

The pseudopotentials used in DFT calculations
may be optimized for various purposes, but should generally
yield consistent and transferable accuracy and results. 
Here, we have used 
some of the pre-defined pseudopotentials\cite{pseudolist}
of the open-source DFT program {\tt DACAPO}.\cite{dacapo}
The values that we find for the lattice constants, for $B_0$, and for $B'_0$,
are collected in 
Table~\ref{table1}.  For reference, the experimental values
are also included, as well as the 
bulk modulii from a Murnaghan and Birch fit along the hydrostatic path. 

We have calculated the structure and bulk modulus for three multiparameter 
systems, as well as for a number of related one-parameter systems:
the two-parameter hcp phase of Co, the three-parameter (one internal)
wurtzite phase of SiC (2H), the five-parameter
(three internal parameters) hexagonal 4H-polytype
of SiC, and the one-parameter bcc and fcc phases of Co, 
the zinc-blende (3C) phase of SiC and the diamond phases
of C and Si.

For the DFT calculations we used the plane-wave {\tt DACAPO} 
code\cite{dacapo} with GGA.
For the calculations of the 2H and 4H-polytype of SiC we used
$8 \times 8\times 8$ 
and $8 \times 8\times 4$ $k$-points, respectively, to describe the 
Brillouin zone.
For all other calculations $10 \times 10\times 10$ $k$-points were used.
A uniform energy cut-off of 400 eV, and a conservative choice 
of fast-Fourier transform (FFT) grid was used. For each evaluation
of the optimal structure we calculated by DFT a number of data points 
for the lattice parameter(s) approximately within 
$\pm 10$\% from the expected optimal value(s) of the lattice parameter(s).
In the one-parameter systems we calculated 20-30 data points, for the 
Co hcp-structure 120 data points, for the 2H-polytype
of SiC 140 data points, and for 4H-SiC we calculated 7500 data points.
The Co calculations were spin-polarized, yielding
realistic values of the spin polarization over the range of
lattice parameter values considered here. 

The possibility of treating the external and internal parameters
collectively is important. For example, the variation of internal 
bond length with pressure might be as important for the
total energy as the change in (external) lattice parameters. 
Often, relaxation of the internal parameters is done with a steepest-descent
(or similar) search minimizing the Hellmann-Feynman forces to a certain
cut-off at fixed lattice parameters.
Although in practice the atomic relaxation will often be a convenient 
way of obtaining the optimal position of the atoms within the unit cell,
the approach has two shortcomings: it introduces a random residual lattice 
strain, which in turn affects the total energy, and further, 
the Hellmann-Feynman forces have a non-trivial dependence
on the pressure acting on the unit cell and therefore a constant cut-off
on the force will not correspond to a constant accuracy of the total energy
with varying pressure.
Thus a  
better accuracy --- and a consistent choice of accuracy --- can be obtained
by treating the lattice and internal parameters on an equal footing.
This is here done for the 2H and the 4H polytype of SiC. 
The results are shown in Table~\ref{table1}.
For the Murnaghan and Birch values of $B_0$ and $B'_0$ we need to
explicitly calculate the hydrostatic path [in $(a,c/a,u)$ and 
$(a,c/a,u_1,u_2,u_3)$ space] before obtaining the fit.
In contrast, we stress that when using (\ref{eq:mpbulk}) and 
(\ref{eq:mpbulkderiv}) there is no need to explicitly calculate the
hydrostatic path. This is here done purely for illustrational purposes.

\section{Conclusion}

In summary, we have presented a new direct algorithm for a combined
determination of structure and bulk moduli $B_0$ and $B'_0$. 
The lattice constants of a multiparameter system are best found in a 
least-squares polynomial fit, as previously noticed for the SiC hexagonal 
polytypes.\cite{kackell}
We show (a) how to exploit this polynomial fit for a direct determination 
of the zero-pressure bulk modulus $B_0$ and its pressure derivative $B'_0$,
avoiding the calculation of the hydrostatic path and the subsequent 
one-dimensional fit to this path. We further show (b) how to consistently 
include internal parameters, 
such as bond lengths or bonding angles, in the formalism along
with the external lattice parameters.  In addition, we
have evaluated these formal results  in explicit cases within our 
approach, based on DFT calculations.

\begin{acknowledgments}
We thank Per Hyldgaard for useful discussions. This project
was partly supported by the EU Human potential research training network 
ATOMCAD under contract number HPRN-CT-1999-00048. Further, E.S.~thanks 
the Swedish Research Council (VR) and the foundations Trygger and
Lundgren for economic support.
\end{acknowledgments}


\begin{thebibliography}{99}
\bibitem{comment} Here we stress that polynomial fitting to the entire, 
multidimensional, free-energy variation is more accurate than the 
frequently-used two-step approach of first finding an explicit expression 
for the hydrostatic path followed by a one-dimensional minimization 
along this path.
\bibitem{pseudolist} The following implementations of pseudopotentials
were used, all part of the \texttt{DACAPO} package (v.~2.4.5): 
\texttt{Co\_us\_gga}, \texttt{csi\_e8ag4}, and \texttt{C\_us\_gga}.
\end{thebibliography}
\end{document}